\documentstyle[12pt,epsf]{article}
\newcommand\pubnumber{SLAC-PUB-8084\\
                        SU-ITP-99/15}
\newcommand\pubdate{March 20, 1999}
\def\Title#1{\begin{center} {\Large #1 } \end{center}}
\def\Author#1{\begin{center}{ \sc #1} \end{center}}
\def\Address#1{\begin{center}{ \it #1} \end{center}}

\def\nsfgrant{\footnote{Work partially supported by NSF grant
PHY-9870115.}}
\def\doeack{\footnote{Work supported by the Department of Energy,
                     contract DE--AC03--76SF00515.}}
\def\Stanford{Physics Department\\
     Stanford University, Stanford, California 94309 USA}
\def\SLAC{Stanford Linear Accelerator Center\\
    Stanford University, Stanford, California 94309 USA}
\newcommand\pubblock{\rightline{\begin{tabular}{l} \pubnumber\\
         \pubdate  \end{tabular}}}
\newenvironment{Abstract}{\begin{quotation} \begin{center}
                       ABSTRACT
     \end{center}\bigskip  }{\end{quotation}}
\def\beq{\begin{equation}}
\def\eeq#1{\label{#1}\end{equation}}
\def\eeqn{\end{equation}}
\def\beqa{\begin{eqnarray}}
\def\eeqa#1{\label{#1}\end{eqnarray}}
\def\eeqan{\end{eqnarray}}
\def\st#1{\scriptstyle {\rm #1}}
\def\CR{\nonumber \\ }
\def\leqn#1{(\ref{#1})}

\def\stacksymbols #1#2#3#4{\def\theguybelow{#2}
    \def\vp{\lower#3pt}
    \def\sp{\baselineskip0pt\lineskip#4pt}
    \mathrel{\mathpalette\intermediary#1}}
\def\intermediary#1#2{\vp\vbox{\sp
     \everycr={}\tabskip0pt
     \halign{$\mathsurround0pt#1\hfil##\hfil$\crcr#2\crcr
              \theguybelow\crcr}}}
\def\gapproxeq{\stacksymbols{>}{\sim}{2.5}{.2}}
\def\lapproxeq{\stacksymbols{<}{\sim}{2.5}{.2}}
\def\Journal#1#2#3#4{{#1} {\bf #2}, #3 (#4)}

\def\NPB{{\em Nucl. Phys.} B}
\def\PLB{{\em Phys. Lett.}  B}
\def\PRL{{\em Phys. Rev. Lett.}}
\def\PRD{{\em Phys. Rev.} D}
\def\PR{{\em Phys. Rev.}}
\def\PRep{{\em Phys. Rep.}}

\def\Dokl{{\em Dokl. Akad. Nauk. SSSR}}
\def\st{\scriptstyle}

\let\bar=\overbar
\def\Dslash{\not{\hbox{\kern-4pt $D$}}}
\def\dslash{\not{\hbox{\kern-2pt $\del$}}}
\def\half{\frac{1}{2}}

\def\del{\partial}

\def\ee{e^+e^-}

\def\msb{{\bar{\ssstyle M \kern -1pt S}}}

\def\eps{\epsilon}
\def\mn{m_N}
\def\mg{m_g}
\def\kkll{|\vec{k}|^2 |\vec{l}|^2}
\def\kl{(\vec{k} \cdot \vec{l})}
\def\etal{{\it et al.}}

\def\eg{{\it e.g.}}

\begin{document}
\begin{titlepage}
\pubblock
\vfill \Title{SN1987A Constraints on Large Compact Dimensions}
\vfill \Author{Schuyler Cullen\nsfgrant} \Address{\Stanford}
\Author {Maxim Perelstein\doeack} \Address{\SLAC} \vfill
\begin{Abstract}
Recently there has been a lot of interest in models in which
gravity becomes strong at the TeV scale. The observed weakness of
gravitational interactions is then explained by the existence of
extra compact dimensions of space, which are accessible to gravity
but not to Standard Model particles. In this letter we consider
graviton emission into these extra dimensions from a hot supernova
core. The phenomenology of SN1987A places strong constraints on
this energy loss mechanism, allowing us to derive a bound on the
fundamental Planck scale. For the case of two extra dimensions we
obtain a very strong bound of M $\gapproxeq$ 50 TeV, which
corresponds to a radius R $\lapproxeq$ 0.3 $\mu$m. While there are
a lot of sources of uncertainty associated with this bound, we
find that pushing it down to the few-TeV range, which could in
principle be probed by ground-based experiments, is disfavored.
For three or more extra dimensions the SN1987A constraints do not
exclude a TeV gravitational scale.
\end{Abstract}
\medskip

\vfill
\end{titlepage}
\def\thefootnote{\fnsymbol{footnote}}
\setcounter{footnote}{0}
Recently, Arkani-Hamed, Dimopoulos and Dvali proposed a novel
solution to the hierarchy problem, which does not rely on either
low-energy supersymmetry or compositeness of the Higgs boson
\cite{ADD1, ADD2, ADD3}. They pointed out that if there exist new
compact spatial dimensions, the fundamental (higher-dimensional)
Planck scale   $M$ could be close to the electroweak scale
$M_{\hbox{$\st{\rm EW}$}}$, thus avoiding the hierarchy. That is,
gravity could become comparable in strength to other interactions
at energies of TeV order. At distances large compared to the size
of the compact dimensions, gravity obeys the 4-dimensional
Newton's Law, with the gravitational constant given by $$ G_N={1
\over 4 \pi} M^{-n-2} R^{-n}, $$ where $n$ is the number of extra
dimensions, and $R$ is their size\footnote{ Throughout this paper,
we will assume for simplicity that the new dimensions are
compactified on a torus of periodicity $2 \pi R$ in each
direction.}. In order to reproduce the measured value $G_N = 6.7
\times 10^{-39}$ GeV$^{-2}$ with the fundamental scale $M \sim$ 1
TeV, we require \beq R \sim 2 \times 10^{31/n-17} \hbox{cm}.
\eeq{R} The most obvious experimental consequence of this scenario
is the violation of Newton's Law at distances of order $R$.
Macroscopic measurments of gravity constrain $R$ to be less than
about a millimeter \cite{Price}. From \leqn{R} we see that for
$n=1$, $R \sim 10^{13}$ cm, so this case is clearly excluded. For
$n \geq 2$, however, this constraint is satisfied: $M=1$ TeV
corresponds to $R=0.68$ mm for $n=2$ and to $R=3.0 \times
10^{-12}$ cm for $n=6$.

Since the Standard Model provides an accurate description of the strong,
electromagnetic and weak processes up to energies of order 100 GeV, the
quarks, leptons and gauge bosons cannot propogate in the extra dimensions.
Therefore they must be localized to a 4-dimensional hypersurface within
the full space-time. While such a localization may be achieved by purely
field-theoretic mechanisms \cite{ADD1}, the most attractive possibility is
provided by D-branes of string theory \cite{ADD2}, since they naturally
appear with certain degrees of freedom confined to them.

If indeed gravity becomes strong at TeV scale, the
higher-dimensional gravitons should have significant couplings to
the Standard Model particles at energies accessible to current and
near-future collider experiments \cite{ADD3}. In particular,
graviton production in $\ee$ and $p\bar{p}$ collisions leads to
events with missing energy\cite{Jim, us}. The current results from
searches for such events at LEP and Tevatron require that $M$ be
higher than 1.2 TeV for $n=2$ and 610 GeV for $n=6$ \cite{us}. A
number of authors have also considered the effects of the virtual
graviton exchange on various observables \cite{Jim, Virtual}, as
well as graviton loop effects on the anomalous magnetic moment of
the muon \cite{Graesser}.

Various astrophysical and cosmological processes can also be used
to put constraints on the model. In particular, the
order-of-magnitude estimates in \cite{ADD3} show that the
agreement of the observed neutrino fluxes from the supernova
SN1987A with the predictions of the stellar collapse models
\cite{Raffelt} requires that the fundamental scale be as high as
$M \sim 30$ TeV for $n=2$. Since this is one of the most stringent
bounds known on $M$, we feel that a more detailed quantitative
study of graviton emission from the supernova is
warranted. In particular, the next-generation collider experiments, as
well as (for the case of two extra dimensions) the upcoming precision
measurements of gravity on short distances, could
probe values of $M$ up to a few TeV \cite{Price, Jim, us}. Therefore it
is interesting to ask whether this range is already ruled out by SN1987A.
We will address this question in this letter.

According to the standard theory of type-II supernovae, most of
the $\sim 10^{53}$ ergs of gravitational binding energy released
during core collapse is carried away by neutrinos. This hypothesis
was essentially confirmed by the measurements of neutrino fluxes
from SN1987A by Kamiokande \cite{Kamio} and IMB
\cite{IMB} collaborations. These measurements allow one to put
powerful constraints on new physics. Indeed, if there is some
novel channel through which the core of the supernova can lose energy, the
luminosity in this channel should be low enough to preserve the
agreement of neutrino observations with theory. This idea was used
to put the strongest experimental upper bounds on the axion mass
\cite{axions}. Here, we will consider the emission of the
higher-dimensional gravitons\footnote{It is likely that,
in realistic models, the extra dimensions will contain scalar, vector,
and even fermion fields that have substantial couplings to matter
\cite{Flavor}. While these particles could also contribute to the energy
loss from the core and strenghten our bound, we do not consider them here,
since this contribution is
model-dependent.} from the core. Once these particles
are produced, they escape into the extra dimensions, carrying
energy away with them. The constraint on the luminosity of this
process can be converted into a bound on the fundamental Planck scale of
the theory, $M$.

During the first few seconds after collapse, the core contains
neutrons, protons, electrons, neutrinos and thermal photons. There
are a number of processes in which higher-dimensional gravitons
can be produced. For the conditions that pertain in the core at
this time (temperatures $T \sim 30-70$ MeV, densities
$\rho \sim (3-10) \times 10^{14}$ g
cm$^{-3}$), the dominant process is nucleon-nucleon
``gravi-strahlung'', \beq N+N \rightarrow N+N+G, \eeq{process} where
$N$ can be a neutron $n$ or a proton $p$, and $G$ is a
higher-dimensional graviton.  Electromagnetic processes in which
gravitons can be emitted, such as $ep \to epG$ and $ee \to eeG$,
are much less important at temperatures of interest than the strong force
mediated reaction \leqn{process}. The processes involving
photons, $\gamma \gamma \to G$, $\gamma e \to eG$ and $\gamma
p \to pG$, are also suppressed, since the photons have
significantly smaller number density than nucleons. In this paper,
we will only consider reaction \leqn{process}.

To model the nucleon-nucleon interaction, we have employed the
one-pion exchange approximation. The nucleon-pion interaction
Lagrangian is \beq {\cal L}_{NN\pi} = -{i g_A \over F_\pi}
\partial_{\mu} \pi^{a} \bar{N} \gamma_5 \gamma^{\mu} \tau^a N,
\eeq{npi} where $N^T=(p, n)$, $\tau^a$ are Pauli matrices, $g_A
\simeq 1.25$, and $F_{\pi} \simeq 180$ MeV. While the validity of
the one-pion exchange potential at nuclear densities is
questionable, we believe it is appropriate to use it for our
present purposes, since factors of order unity in the graviton
emission rate will not have much effect on our conclusions.

%%%%%%%%%%%%%%%%%%%%%%%%%%%%%%%%%%%%%%%%%%%%%%%%%%%%%%%%%%%%%%%%%%%%%%
\begin{figure}
\centerline{\epsfysize=3.00truein \epsfbox{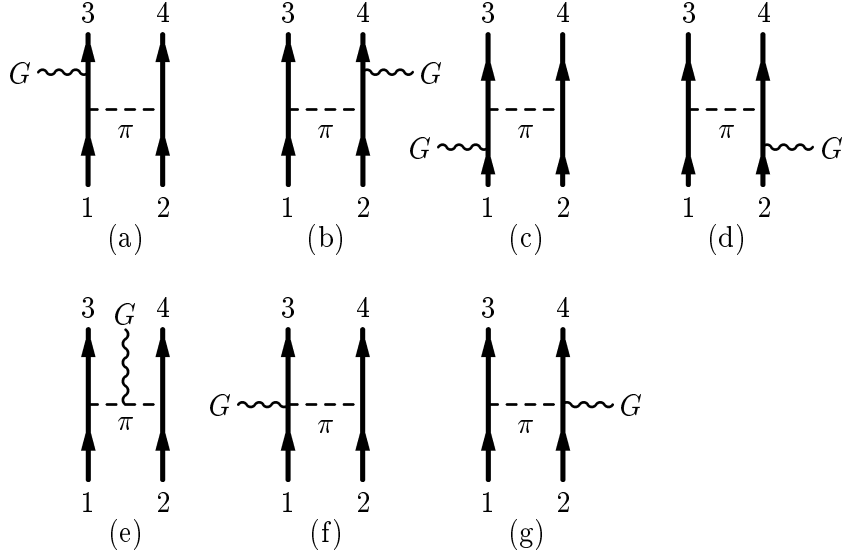}}
\vskip -0.4 cm
 \caption{Diagrams that contribute to the gravi-strahlung process
in nucleon-nucleon collision in the one-pion exchange approximation.}
\label{diagrams}
\end{figure}
%%%%%%%%%%%%%%%%%%%%%%%%%%%%%%%%%%%%%%%%%%%%%%%%%%%%%%%%%%%%%%%%%%%%%%
%

From the four-dimensional point of view, a higher-dimensional
graviton appears as a tower of ``Kaluza-Klein modes'', massive
particles whose mass is determined by their momentum in the extra
dimensions and is quantized in units of $1/R$. The typical
energies at which gravitons are emitted in the core are much lower
than the fundamental scale; in this limit, only spin-2 modes
(which we will refer to as ``gravitons'', $h^{\mu\nu}$ ) and
spin-0 modes (``dilatons'', $\phi$) couple to the Standard Model
particles\footnote{This fact is obvious from the effective field
theory point of view, developed in \cite{Sundrum}.}. The
Lagrangian is \cite{Han}: \beq {\cal L} = -{\kappa \over 2}
\sum_{\vec{j}} \left (h^{\mu\nu,\vec{j}} T_{\mu\nu} + \omega
\phi^{\vec{j}} T^{\mu}_{\,\mu} \right), \eeq{Han} where
$T_{\mu\nu}$ is the conserved energy-momentum tensor of the
matter, $\kappa=\sqrt{32 \pi G_N}$, $\omega = \sqrt{{2 \over
3(n+2)}}$, and the sum is over the various Kaluza-Klein (KK)
modes, labelled by their momentum in the compact dimensions
$\vec{j}$. Note that each KK mode couples to matter with the usual
four-dimensional gravitational strength. However, for $n \le 4$,
the number of these modes that can be emitted at the core
temperatures is very large: the mass splitting between the modes
ranges from 10$^{-4}$ eV for $n=2$ to about 10 keV for $n=4$. This
enormous multiplicity leads to enhancement of their effects, and
allows us to put strong constraints on the model. For $n>4$, the
mass splitting becomes comparable to the temperatures in the core,
so that only the first few modes can be emitted. In this case, the
bound we obtain is very weak, and we will not discuss it here.
Note also that once a particular mode is emitted, it practically
does not interact with matter. Thus, the gravitons, unlike the
axions and neutrinos, cannot be trapped in the core. We emphasize
that the Lagrangian \leqn{Han} is completely independent of the
details of the physics at the fundamental scale $M$. Therefore,
our results will provide a model-independent bound on this scale.

Given the Lagrangians \leqn{npi} and \leqn{Han}, it is a simple
matter to evaluate the matrix elements for each of the processes
in \leqn{process}. There are 14 diagrams to evaluate, the seven
``direct'' diagrams shown in Figure~\ref{diagrams}, and seven
``exchange'' diagrams, with legs 3 and 4 interchanged. Since the
temperatures in the core are of order 30-70 MeV, the nucleons are
nonrelativistic. Naively, one would expect the diagrams (a)-(d)
(and the corresponding exchange diagrams) to dominate in this
regime. However, the leading contributions cancel due to energy
and momentum conservation, and all the diagrams have to be
included in the calculation. We have also made the approximation
that $m_\pi^2 \ll 3 m_N T$; this is satisfied for $T \gg 6$ MeV.
For the emission of a single Kaluza-Klein graviton mode $\vec{j}$
we find \vspace{-0.75cm}
\begin{center}
\beqa \nonumber \sum_{\hbox{$\st{\rm spin}$}} |{\cal
M}^{\vec{j}}(nn \rightarrow nnG)|^2 = \sum_{\hbox{$\st{\rm
spin}$}} |{\cal M}^{\vec{j}}(pp \rightarrow ppG)|^2 = {\cal
A}^{\vec{j}} + {\cal B}^{\vec{j}} - 2 {\cal C}^{\vec{j}} ,
\eeqa{nnpp} \vspace{-0.5cm} \beqa \sum_{\hbox{$\st{\rm spin}$}}
|{\cal M}^{\vec{j}}(np \rightarrow npG)|^2 = {\cal A}^{\vec{j}} +
4 {\cal B}^{\vec{j}} + 4 {\cal C}^{\vec{j}}; \eeqa{nnpp}
\end{center}
\vspace{-0.25cm} where ${\cal A}^{\vec{j}}$ and ${\cal
B}^{\vec{j}}$ are the contributions of the direct and exchange
diagrams, respectively, and ${\cal C}^{\vec{j}}$ is due to the
interference between the two types of diagrams. A minus sign
appears in the first line due to exchange of identical fermions in
the final state. In $np$ scattering, the contributions of exchange
diagrams are enhanced because they involve charged pions, which
couple to nucleons more strongly than neutral pions by a factor of
$\sqrt{2}$. Explicitly, we find \beqa \nonumber {\cal
A}^{\vec{j}}&=&{\cal B}^{\vec{j}}
\\ \nonumber &=&{1024 \pi \over 45} {\mn^4g_A^4 G_N \over F_\pi^4}
\Biggl[ 27 + 26 {\mn^2 \mg^2 \over \kl^2} + 7 {\mn^4 \mg^4 \over
\kl^4} + {\kkll \over \kl^2} \left( 19 + 22 {\mn^2 \mg^2 \over
\kl^2} + 4{\mn^4 \mg^4 \over \kl^4} \right) \Biggr], \\ \nonumber
\\  \nonumber {\cal C}^{\vec{j}} &=& \hskip -4.5pt -{512 \pi \over
45} {\mn^4g_A^4  G_N \over F_\pi^4} \Biggl[ 27 + 26 {\mn^2 \mg^2
\over \kl^2} + 7 {\mn^4 \mg^4 \over \kl^4} - 5 {\kl^2 \over \kkll}
\left( 7 + 6 {\mn^2 \mg^2 \over \kl^2} + 2 {\mn^4 \mg^4 \over
\kl^4} \right)  \\  & & \hskip 1.2in + \left( {\kkll \over \kl^2}
- 4 {\kl^4 \over |\vec{k}|^4 |\vec{l}|^4} \right)
  \left( 19 + 22 {\mn^2 \mg^2 \over \kl^2} + 4 {\mn^4 \mg^4 \over \kl^4}
\right) \Biggr]; \eeqa{Msqared}
where $\vec{k}=\vec{p_1}-\vec{p_3}$,
$\vec{l}=\vec{p_1}-\vec{p_4}$, and $\mg$ is the four-dimensional
mass of the emitted graviton mode. These expressions have been
averaged over the direction of the graviton. We have also
evaluated the matrix elements for the dilaton emission, but since
their contribution to the energy loss rate turns out to be
negligible we do not present them here.

The energy loss rate for the nucleon-nucleon gravi-strahlung is
given by the following phase-space integral:
\beq
\dot\eps=
\sum_{\vec{j}} \int d\Pi_1 d\Pi_2 d\Pi_3 d\Pi_4 d\Pi_g S | {\cal
M}^{\vec{j}} |^2 (2 \pi)^4 \delta^4 (p_1+p_2-p_3-p_4-p_g) E_g f_1
f_2 (1-f_3) (1-f_4),
\eeq{integral}
where $d \Pi_i=d^3p_i/(2\pi)^3 2 E_i$, the labels $i$ denote the
incoming ($i=1,2$) and outgoing ($i=3,4$) nucleons, the label $g$
denotes the higher-dimensional graviton, $S$ is the symmetry factor for
identical particles in the initial and final states ($S=1/4$ for
$nn \rightarrow nn$, $pp \rightarrow pp$, and $S=1$ for $np
\rightarrow np$), and $f_i=[\exp(E_i/T-\mu_i/T)+1]^{-1}$ are the
distribution functions of the nucleons. Since the typical energies
in our process are much higher than the splitting of the KK modes,
we can replace the sum over these modes by an integral over the
four-dimensional mass, according to \beq
   \sum_{\vec{j}} \rightarrow   R^n  \int d^n m =
  \half \Omega_n R^n \,  \int\,  (m^2)^{(n-2)/2} d\,  m^2 =
    {\Omega_n\over 8 \pi} M^{-(n+2)} G_N^{-1} \int\,  (m^2)^{(n-2)/2} d\,
      m^2 ,
\eeq{mps}
where $\Omega_n$ is the surface area of the unit sphere in $n$ dimensions.

We will assume that the nucleons in the core are non-degenerate;
in this limit, $f_i=\exp(\mu_i/T-E_i/T)$ and we can neglect the
``blocking factors'', $1-f_3$ and $1-f_4$. Integrating the matrix
elements \leqn{nnpp} over the phase-space in \leqn{integral} and
over the four-dimensional mass of the emitted graviton, we get
\beqa \dot{\eps} &=& 1.7 \times 10^{17} \,\, \hbox{erg
g}^{-1}\hbox{s}^{-1} \left( X_n^2 + X_p^2 + 7.0 X_nX_p \right)
\rho_{14} T_{\hbox{$\st{\rm MeV}$}}^{5.5} M_{\hbox{$\st{\rm
TeV}$}}^{-4}, \hskip0.5cm n=2; \CR \dot{\eps} &=& 9.4 \times
10^{11} \,\, \hbox{erg g}^{-1}\hbox{s}^{-1} \left( X_n^2 + X_p^2 +
7.8 X_nX_p \right) \rho_{14} T_{\hbox{$\st{\rm MeV}$}}^{6.5}
M_{\hbox{$\st{\rm TeV}$}}^{-5}, \hskip0.5cm n=3; \CR \dot{\eps}
&=& 5.9 \times 10^{6} \,\,\,\, \hbox{erg g}^{-1}\hbox{s}^{-1}
\left( X_n^2 + X_p^2 + 8.8 X_nX_p \right) \rho_{14}
T_{\hbox{$\st{\rm MeV}$}}^{7.5} M_{\hbox{$\st{\rm TeV}$}}^{-6},
\hskip0.5cm n=4; \eeqa{rates} where $X_p$ and $X_n=1-X_p$ are the
proton and neutron fractions in the core, $\rho_{14}=\rho/(10^{14}
\,\hbox{g cm}^{-3})$, $T_{\hbox{$\st{\rm MeV}$}}= T/(1 \,
\hbox{MeV})$, and $M_{\hbox{$\st{\rm TeV}$}}=M/(1 \, \hbox{TeV})$.

In order to obtain reliable bounds on the fundamental Planck scale
$M$, one would have to incorporate the luminosities \leqn{rates}
into a numerical code for a protoneutron star evolution, and
calculate the expected neutrino fluxes for various values of $M$.
For the purposes of this paper, however, we will adopt a simple
analytic criterion, suggested by Raffelt (see p. 504 of
\cite{Raffelt}.) Namely, we will require that the energy loss rate
to gravitons, evaluated at typical core conditions, does not
exceed $10^{19}$ erg g$^{-1}$ s$^{-1}$. Using \leqn{rates} then
yields, for the case of two extra dimensions, \beq M > 0.36 \,\,
\hbox{TeV} (X_n^2+X_p^2+7.0X_nX_p)^{0.25} \, \rho_{14}^{0.25}
\,T_{\hbox{$\st{\rm MeV}$}}^{1.375}, \hskip2cm n=2. \eeq{bound}
This bound depends rather weakly on the proton fraction, core
density, and the exact value of the maximum luminosity we allow.
On the other hand, its temperature dependence is strong.
Therefore, the main uncertainty comes from the lack of precise
knowledge of temperatures in the core: values quoted in the
literature range from 30 to 70 MeV. For a numerical estimate, we
have assumed, conservatively, $T=30$ MeV, $\rho=3 \times 10^{14}$
g cm$^{-3}$, and $X_p=0$. This choice of parameters yields \beq M
\, \gapproxeq \, 50 \,\, \hbox{TeV}, \hskip2in n=2.
\eeq{numbound2} This corresponds to an extra dimension size of
\beq R \, \lapproxeq 3 \times \, 10^{-4} \,\, \hbox{mm},
\hskip1.65in n=2. \eeq{Rbound2} These values of $M$ and $R$ are
well beyond the reach of both current and near-future ground-based
experiments. We note that for higher realistic values of
temperatures and densities in the core the bound obtained from
\leqn{bound} can be as high as 250 TeV.

In the course of our calculation, we have made a number of
simplifying approximations. In particular, we have not considered
multiple-scattering effects, which suppress emission of gravitons
with energies below the nucleon-nucleon collision rate in the
medium\footnote {This phenomenon is analogous to the
Landau-Pomeranchuk-Migdal effect \cite{LPM} in the usual photon
bremstrahlung.}. The deviations of the nucleon mass and
nucleon-pion coupling from their vaccuum values in the dense
medium of the core were also neglected. While at the core
temperatures the nucleons are neither strongly degenerate nor
non-degenerate, we have not included the degeneracy effects, such
as the Pauli blocking factors for the final-state nucleons. One
may also worry about the inadequacies of the one-pion exchange
approximation at short distances, \eg ~the effects of pion loop
diagrams, other meson exchanges, and the hard core of the
nucleon-nucleon potential. Finally, we have not considered the
possibility that novel phases of nuclear matter (pion condensates,
quark-gluon plasma, superfluidity, etc.) may form in certain
regions of the core.

To make a crude estimate of the uncertainties introduced by
neglecting the above effects, we used the results for the very
similar process of nucleon-nucleon axion bremstrahlung, which has
been studied in great detail. For this process, it was shown that
a simple calculation based on the same approximations we have made
here gives a good first estimate of the bound on axion-nucleon
coupling, $g_a \lapproxeq 10^{-10}$ \cite{Raffelt}. This is only
about a factor of 3 less restrictive than the generally accepted
bound, corresponding to about a factor of 10 uncertainty in the
axion luminosity. We expect a similar result to hold for the
process considered here, even though a much more careful study is
of course necessary to confirm it. If this is the case, our bound
on $M$ is quite robust, since changing the graviton luminosity by
a factor of 10 only changes the bound by $10^{0.25} \simeq 1.8$.
Note that in order to push this bound down to the range of a few
TeV, which in principle could be probed by both collider and
gravitational experiments, our naive calculation would have to
overestimate the luminosity by more than 3 orders of magnitude. We
feel therefore that it is fair to say that this range is
disfavored, and likely ruled out, by SN1987A.

We have repeated our analysis for $n=3$ and $n=4$. With the same
parameter choices as before, we obtain the bounds: \beqa M \,
&\gapproxeq& \, 4 \,\, \hbox{TeV}, \hskip1.5cm R \, \lapproxeq \,
4 \times 10^{-7} \,\, \hbox{mm}, \hskip2cm n=3; \CR M \,
&\gapproxeq& \, 1 \,\, \hbox{TeV}, \hskip1.5cm R \, \lapproxeq \,
2 \times 10^{-8} \,\, \hbox{mm}, \hskip2cm n=4. \eeqa{bounds34}
While these bounds are quite strong, our conclusions in this case
are much more optimistic. Given the uncertainties of our analysis,
models with large extra dimensions and values of the fundamental
scale which could be probed by near-future collider experiments
are certainly not excluded for $n\ge3$.

In summary, we have found that graviton emission from SN1987A puts
very strong constraints on models with large extra dimensions in
the case $n=2$. In this case, for a conservative choice of the
core parameters we arrive at a bound on the fundamental Planck
scale $M\gapproxeq$ 50 TeV, which corresponds to a radius R
$\lapproxeq$ 0.3 $\mu$m. Even though taking into account various
uncertainties encountered in our calculation can weaken this
bound, it is unlikely that it can be pushed down to the
phenomenologically interesting range of a few TeV. For $n=3$ and
4, we find that the fundamental scale has to be higher than about
$4$ and $1$ TeV, respectively.

We are grateful to Nima Arkani-Hamed, Savas Dimopoulos, Lance
Dixon, Michael Peskin, Georg Raffelt, Scott Thomas and Robert
Wagoner for useful discussions. M. P. was supported by the
Department of Energy under contract DE-AC03-76SF00515. S. C. was
partially supported by NSF grant PHY-9870115.

%%%%%%%%%%%%%%%%%%%%%%%%%%%%%%%%%%%%%%%%%%%%%%%%%%%%%%%%%%%%%%%%%%%%%%%%%


\newpage
\begin{thebibliography}{99}

\bibitem{ADD1}
     N. Arkani-Hamed, S. Dimopoulos, and G. Dvali,
             \Journal{\PLB}{429}{263}{1998}.

\bibitem{ADD2}

      I. Antoniadis, N. Arkani-Hamed, S. Dimopoulos, and G. Dvali,
        \Journal{\PLB}{436}{257}{1998}.

\bibitem{ADD3}
     N. Arkani-Hamed, S. Dimopoulos, and G. Dvali,
         \Journal{\PRD}{59}{086004}{1999}.

\bibitem{Price}
   Macroscopic bounds on modifications of gravity have recently been
   reviewed by J. C. Long, H. W. Chan, and J. C. Price,
    \Journal{\NPB}{539}{23}{1999}.

\bibitem{Jim}
   G. F. Giudice, R. Rattazzi, and J. D. Wells,
           \Journal{\NPB}{544}{3}{1999}.

\bibitem{us}
   E. A. Mirabelli, M. Perelstein, and M. E. Peskin,
          \Journal{\PRL}{82}{2236}{1999}.

\bibitem{Virtual}
   J. L. Hewett, hep-ph/9811356;
   P. Mathews, S. Raychaudhuri, and K. Sridhar, hep-ph/9812486;
   T. G. Rizzo, hep-ph/9901209;
   K. Agashe, N. G. Deshpande, hep-ph/9902263.

\bibitem{Graesser}
   M. L. Graesser, hep-ph/9902310.

\bibitem{Kamio}
   K. Hirata \etal, \Journal{\PRL}{58}{1490}{1987}.

\bibitem{IMB}
   R. M. Bionta \etal, \Journal{\PRL}{58}{1494}{1987}.

\bibitem{Raffelt}
   G. G. Raffelt, {\it Stars as Laboratories for Fundamental
   Physics} (Chicago, USA: Univ. Pr., 1996).

\bibitem{axions}
   Originally, the SN1987A bounds on the axion mass were discussed in
   R. Mayle \etal, \Journal{\PLB}{203}{188}{1988};
   G. G. Raffelt and D. Seckel, \Journal{\PRL}{60}{1793}{1988};
   M. S. Turner, \Journal{\PRL}{60}{1797}{1988}.
   Subsequently, a lot of work has been done on refining these bounds.
   For reviews, see G. G. Raffelt, \Journal{\PRep}{198}{1}{1990},
   and Ref. \cite{Raffelt}.

\bibitem{Flavor}
  N. Arkani-Hamed, S. Dimopoulos, hep-ph/9811353.

\bibitem{Sundrum}
   R. Sundrum, \Journal{\PRD}{59}{085009}{1999}.

\bibitem{Han}
   T. Han, J. D. Lykken, and R.-J. Zhang, hep-ph/9811350.

\bibitem{LPM}
   L. D. Landau and I. Pomeranchuk,
   \Journal{\Dokl}{92}{535}{1953}, and  {\bf 92}, 735 (1953);
   A. B. Migdal, \Journal{\PR}{103}{1811}{1956}.

\end{thebibliography}
\end{document}